\documentclass[a4paper,10pt]{article}
\usepackage[dvips]{graphicx}
\usepackage{amssymb,amsmath}
\numberwithin{equation}{section}

\begin{document}
\title{\bf{Integrable geometric flows   of   interacting curves/surfaces,   multilayer  spin systems  and the vector nonlinear Schr\"odinger equation}}
\author{ Akbota  Myrzakul \footnote{Email: akbota.myrzakul@gmail.com} and    Ratbay Myrzakulov\footnote{Email: rmyrzakulov@gmail.com} \\ \textit{Eurasian International Center for Theoretical Physics and  Department of General } \\ \textit{ $\&$  Theoretical Physics, Eurasian National University, Astana 010008, Kazakhstan}}


\date{}
\maketitle
\begin{abstract}
In this paper, we  study    integrable multilayer spin systems, namely, the multilayer M-LIII equation. We investigate their  relation  with the geometric flows of interacting curves and surfaces in some space $R^{n}$. Then we present their  the Lakshmanan equivalent counterparts. We show that these equivalent counterparts are, in fact, the vector nonlinear Schr\"odinger equation (NLSE). It is well-known that the vector NLSE  is equivalent to the $\Gamma$-spin system. Also,  we have presented the transformations which give the relation between solutions of the $\Gamma$-spin system and the multilayer  M-LIII equation. It is interesting to note that the integrable multilayer M-LIII equation contains constant magnetic field ${\bf H}$. It seems that this constant magnetic vector plays an important role in theory of  "integrable multilayer spin system" and in nonlinear dynamics of magnetic systems. Finally, we present some classes of integrable models of interacting vortices. 
\end{abstract}
\vspace{2cm}
\section{Introduction}
There is an interesting geometric relationship between geometric flows of curves and surfaces and integrable nonlinear differential equations \cite{Gut}-\cite{Kostov}.
In the pionering works of nineteenth-century geometers  were already established main connections between the theory of ordinary and partial differential equations and differential geometry of curves and surfaces. In soliton theory, the well-known Darboux and B\"acklund transformations have origins in the relationship between solutions of the sine-Gordon equation and pseudospherical surfaces. In this and previous our papers, we launch the programm to study  integrable flows  of curves and surfaces related with the vector (multi-component) nonlinear Schr\"odinger equation (NLSE) and with multilayer spin systems. Under "curve flows" we usually mean the following equations
\begin{eqnarray}
{\bf e}_{ix}={\bf C}\wedge {\bf e}_{i}, \quad {\bf e}_{it}={\bf D}\wedge {\bf e}_{i},\label{3.23}
\end{eqnarray}
where
\begin{eqnarray}
{\bf C}=\tau{\bf  e}_{1}+\kappa{\bf e}_{3}, \quad {\bf D}=\omega_{1}{\bf e}_{1}+\omega_{2}{\bf e}_{2}+\omega_{3}{\bf e}_{3}.\label{3.23}
\end{eqnarray}
Similarly, we can write the geometric  flows of immersed surfaces in some Euclidean space. They are given by the following Gauss-Weingarden equations
\begin{eqnarray}
\chi_{x}=N\chi,  \quad \chi_{t}=M\chi.\label{3.23}
\end{eqnarray}
Integrability conditions of these equations are given by the following Gauss-Mainardi-Codazzi equation 
\begin{eqnarray}
N_{t}-M_{x}+[N,M]=0.\label{3.23}
\end{eqnarray}

The aim of the present work is to provide geometric formulations of multilayer spin systems. We will demonstrate that the multilayer M-LIII equation arises from the flow of interacting curves. The Lakshmanan equivalent counterparts of the multilayer M-LIII equations are derived from inelastic curve flows in the Euclidean space.  Several new results are obtained by considering different classes of curves and surfaces. 

The outline of this paper is as follows.  In the next two sections we give a brief review of the vector NLSE and the M-LIII equation. In Sections 4-7, we consider the multilayer M-LIII equations and their relations with the vector NLSE. In Section 8, we  provide a Hamilton structure of the multilayer M-LIII equation. The $\Gamma$-spin system which is the gauge equivalent counterpart of the vector NLSE is presented. The relation between solutions of the multilayer M-LIII equation and the $\Gamma$-spin system is considered in Section 10. The geometric flows of curves and surfaces were studied in Sections 11 and 12. In Section 13 we present an integrable filament equations of interacting vortices. At last, in  Section 14, we give  our conclusions. 
\section{Brief review of the vector NLSE}

In this section we recall some general facts about the vector NLS equation. The vector NLS equation  has many physical significant applications such as modeling crossing sea waves and in   propagation of  elliptically birefringent optical fibers.  
\subsection{The equation}
The vector NLSE reads as 
\begin{eqnarray}
iq_{1t}+q_{1xx}-vq_{1}&=&0,\label{3.21}\\
 iq_{2t}+q_{2xx}-vq_{2}&=&0,\label{3.23}\\
 \vdots \\
 iq_{Nt}+q_{Nxx}-vq_{N}&=&0,\label{3.23}
\end{eqnarray}
where
\begin{eqnarray}
v+2(|q_{1}|^{2}+|q_{2}|^{2}+\cdot\cdot\cdot+|q_{N}|^{2})=0.\label{3.23}
\end{eqnarray}
The Manakov system is the particular case of the vector NLSE when $N=2$. The Manakov system we studied in \cite{akbota1}-\cite{akbota7}. 
\subsection{Lax representation}

The vector NLS   equation (2.1)-(2.3) is integrable by the inverse scattering method. Its Lax representation (LR) reads as 
\begin{eqnarray}
\Phi_{x}&=&U\Phi,\label{2.1}\\
\Phi_{t}&=&V\Phi, \label{2.2} 
\end{eqnarray}
where $\Phi=(\phi_{1},\phi_{2}, \phi_{3})$ and 
\begin{eqnarray}
U=-i\lambda \Sigma+U_{0}, \quad V=-2i\lambda^{2}\Sigma+2\lambda U_{0}+V_{0}. \label{2.2} 
\end{eqnarray}
Here
\begin{eqnarray}
\Sigma =
\left ( \begin{array}{cccc}
1   & 0     & 0 &0\\
0 & -1    & 0 &0 \\
0   & 0 & -1&0\\
0   & 0 &0&-1\\
\end{array} \right),\quad U_{0} =
\left ( \begin{array}{cccc}
0       & q_{1}  & q_{2} &q_{3}\\
-\bar{q}_{1} & 0      & 0 &0\\
-\bar{q}_{2}  & 0 & 0&0\\
-\bar{q}_{3} & 0      & 0 &0\\

\end{array} \right), 
\end{eqnarray}
\begin{eqnarray}
  V_{0} =
i\left ( \begin{array}{cccc}
|q_{1}|^{2}+|q_{2}|^{2} +|q_{3}|^{2}      & q_{1x}  & q_{2x} &q_{3x}\\
\bar{q}_{1x} & - |q_{1}|^{2}     & -\bar{q}_{1}q_{2} &-\bar{q}_{1}q_{3}\\
\bar{q}_{2x}  & -\bar{q}_{2}q_{1} & -|q_{2}|^{2}&-\bar{q}_{2}q_{3}\\
\bar{q}_{3x} & - \bar{q}_{3}q_{1}     & -\bar{q}_{3}q_{2} &-|q_{3}|^{2}\\
\end{array}\right).\label{2.2} 
\end{eqnarray}

\section{Brief review of the 1-layer  M-LIII equation}
 
\subsection{The  equation}
\subsubsection{The general form} 
For the pedagogical reason, in this section, we present some informations on the M-LIII equation for the 1-layer case. Consider the spin vector ${\bf A}=(A_{1}, A_{2}, A_{3})$, where ${\bf A}^{2}=1$. Let this spin vector obey  the  1-layer   M-LIII equation which reads as
 \begin{eqnarray}
{\bf A}_{t}+{\bf A}\wedge {\bf A}_{xx}+u_{1}{\bf A}_{x}+{\bf F}=0, \label{2.2} 
\end{eqnarray}
where $u_{1}(x,t, A_{j}, A_{jx})$ is the real function (potential), ${\bf F}$ is some vector function.  The matrix form of the M-LIII equation looks like
\begin{eqnarray}
iA_{t}+\frac{1}{2}[A,A_{xx}]+iu_{1}A_{x}+F&=&0,\label{2.2} 
\end{eqnarray}
where
\begin{eqnarray}
A=\begin{pmatrix} A_{3}&A^{-}\\ 
A^{+}&-A_{3}\end{pmatrix}, \quad A^{2}=I=diag(1, 1), \quad A^{\pm}=A_{1}\pm i A_{2}. \label{2.2} 
\end{eqnarray}
\begin{eqnarray} F=\begin{pmatrix} F_{3}&F^{-}\\ 
F^{+}&-F_{3}\end{pmatrix}, \quad F^{\pm}=F_{1}\pm iF_{2}. \label{2.2} 
\end{eqnarray}

\subsubsection{The vector form} 
Sometime we consider the following particular case of the M-LIII equation
\begin{eqnarray}
{\bf A}_{t}+{\bf A}\wedge {\bf A}_{xx}+u_{1}{\bf A}_{x}+v_{1}{\bf H}\wedge {\bf A}=0, \label{2.2} 
\end{eqnarray}
where $v_{1}(x,t, A_{j}, A_{jx})$ is the real function (potential), ${\bf H}=(0,0,1)$ is the constant  magnetic field. 

\subsubsection{The matrix  form} 
The matrix form of this equation has the form
\begin{eqnarray}
iA_{t}+\frac{1}{2}[A,A_{xx}]+iu_{1}A_{x}+v_{1}[\sigma_{3},A]&=&0.\label{2.2} 
\end{eqnarray}

\subsubsection{The w -- form} 
We now introduce a new   complex function $w$  as
\begin{eqnarray}
w=\frac{A^{+}}{1+A_{3}}.\label{2.16}
\end{eqnarray}
Then the M-LIII equation takes the form
 \begin{eqnarray}
iw_{t}-w_{xx}+\frac{2{\bar w}w_{x}^{2}}{1+|w|^{2}}&=&F^{\prime},\label{2.16}
\end{eqnarray}
where $F^{\prime}$  is  some complex function of the form
\begin{eqnarray}
F^{\prime}&=&F^{\prime}(w,w_{x}, {\bar w}, {\bar w}_{x}, u, v,).\label{2.16}
\end{eqnarray}
It is the w-form of the M-LIII equation.

\subsubsection{The r -- form} 
Let us introduce the new function ${\bf R}$ as
\begin{eqnarray}
{\bf R}=\partial^{-1}_{x}{\bf S}.\label{2.2} 
\end{eqnarray}
Then the r-form of the M-LIII equation reads as
\begin{eqnarray}\label{S-Flow}
{\bf R}_{t}={\bf R}_{x}\wedge {\bf R}_{xx}+u{\bf R}_{x}+2v{\bf H}\wedge {\bf R}+{\bf L},
\end{eqnarray}
where 
\begin{eqnarray}\label{S-Flow}
{\bf L}=-\partial_{x}^{-1}[u_{x}{\bf R}_{x}+2v_{x}{\bf H}\wedge {\bf R}].
\end{eqnarray} 
\subsection{The Lakshmanan equivalent counterpart}

Let us find the Lakshmanan equivalent counterpart of the 1-layer  M-LIII equation (2.1)-(2.2).  To do that, consider 3-dimensional curve in  $R^{3}$. This curve is  given by the  following  vectors ${\bf e}_{k}$. These vectors satisfy the following equations 
\begin{eqnarray}
\left ( \begin{array}{ccc}
{\bf  e}_{1} \\
{\bf  e}_{2} \\
{\bf  e}_{3}
\end{array} \right)_{x} = C
\left ( \begin{array}{ccc}
{\bf  e}_{1} \\
{\bf  e}_{2} \\
{\bf  e}_{3}
\end{array} \right),\quad
\left ( \begin{array}{ccc}
{\bf  e}_{1} \\
{\bf  e}_{2} \\
{\bf  e}_{3}
\end{array} \right)_{t} = D
\left ( \begin{array}{ccc}
{\bf  e}_{1} \\
{\bf  e}_{2} \\
{\bf  e}_{3}
\end{array} \right). \label{2.1} 
\end{eqnarray}
Here ${\bf e}_{1}, {\bf e}_{2}$ and ${\bf e}_{3}$ are the unit tangent, normal 
and binormal vectors  to the  curve,  $x$ is its arclength 
parametrising the curve. The matrices $C$ and $G$ have the forms
\begin{eqnarray}
C =
\left ( \begin{array}{ccc}
0   & k_{1}     & 0 \\
-k_{1} & 0     & \tau_{1}  \\
0   & -\tau_{1} & 0
\end{array} \right) ,\quad
G =
\left ( \begin{array}{ccc}
0       & \omega_{3}  & -\omega_{2} \\
-\omega_{3} & 0      & \omega_{1} \\
\omega_{2}  & -\omega_{1} & 0
\end{array} \right).\label{2.2} 
\end{eqnarray}
The   curvature and torsion of the  curve  are given  by the following formulas
\begin{eqnarray}
k_{1}= \sqrt{{\bf e}_{1x}^{2}},\quad \tau_{1}= \frac{{\bf e}_{1}\cdot ({\bf e}_{1x} \wedge {\bf e}_{1xx})}{{\bf e}_{1x}^{2}}.        \label{2.3}
\end{eqnarray}
The compatibility condition of the equations (3.7) is given by
\begin{eqnarray}
C_t - G_x + [C, G] &=& 0,\label{2.4} 
\end{eqnarray}
or in elements   
 \begin{eqnarray}
k_{1t}    & = & \omega_{3x} + \tau_{1} \omega_2, \label{2.5} \\ 
\tau_{1t}      & = & \omega_{1x} - k_{1}\omega_2, \\ \label{2.6} 
\omega_{2x} & = & \tau_{1} \omega_3-k_{1} \omega_1. \label{2.7} 
\end{eqnarray}

Now we do the following identifications:
 \begin{eqnarray}
{\bf A}\equiv {\bf e}_{1},   \quad {\bf F}=F_{1}{\bf e}_{1}+F_{2}{\bf e}_{2}+F_{3}{\bf e}_{3}. \label{2.7} 
\end{eqnarray}
Then we have 
\begin{eqnarray}
k_{1}^{2} & = & {\bf A}_{x}^{2},\\ 
\tau_{1}&=&  \frac{{\bf A}\cdot ({\bf A}_{x} \wedge {\bf A}_{xx})}{{\bf A}_{x}^{2}}, \label{2.3}
\end{eqnarray}
and
\begin{eqnarray}
\omega_{1} & = & -\frac{k_{1xx}+F_{2}\tau_{1}+F_{3x}}{k_{1}}+(\tau_{1}-u_{1})\tau_{1},\\ 
\omega_{2}&=& k_{1x}+F_{3}, \\
\omega_{3} & = &k_{1}(\tau_{1}-u_{1})-F_{2},       \label{2.3}
\end{eqnarray}
with $F_{1}=E_{1}=0$.
The  equations for $k_{1}$ and $\tau_{1}$ reads as
 \begin{eqnarray}
k_{1t}&=&2k_{1x}\tau_{1}+k_{1}\tau_{1x}-(u_{1}k_{1})_{x}-F_{2x}+F_{3}\tau_{1}, \label{2.5} \\ 
\tau_{1t}&=&\left[-\frac{k_{1xx}+F_{2}\tau_{1}+F_{3x}}{k_{1}}+(\tau_{1}-u_{1})\tau_{1}-\frac{1}{2}k_{1}^{2}\right]_{x}-F_{3}k_{1}.  \label{2.7} 
\end{eqnarray}
Next we introduce a new  complex function as 
\begin{eqnarray}
q_{1}=\frac{\kappa_{1}}{2}e^{-i\partial^{-1}_{x}\tau_{1}}. \label{2.2} 
\end{eqnarray}
This function satisfies the following equation
\begin{eqnarray}
iq_{1t}+q_{1xx}+2|q_{1}|^{2}q_{1}+...=0.\label{3.23}
\end{eqnarray}
It is the desired Lakshmanan equivalent counterpart of the M-LIII equation (3.5). If $u_{1}=v_{1}=0$, it turns to the NLSE
 \begin{eqnarray}
iq_{1t}+q_{1xx}+2|q_{1}|^{2}q_{1}=0.\label{3.23}
\end{eqnarray}
\section{The 2-layer  M-LIII equation}
 In  this section we consider  the 2-layer  M-LIII equation. As shown in \cite{akbota1}-\cite{akbota7}, the  2-layer  M-LIII equation is integrable by the inverse scattering transform (IST).
 
\subsection{The  equation }
 In this paper, we consider two  spin vectors ${\bf A}=(A_{1}, A_{2}, A_{3})$ and  ${\bf B}=(B_{1}, B_{2}, B_{3})$, where ${\bf A}^{2}={\bf B}^{2}=1$. Let these spin vectors satisfy the  following 2-layer M-LIII equation or the coupled  M-LIII equation 
\begin{eqnarray}
{\bf A}_{t}+{\bf A}\wedge {\bf A}_{xx}+u_{1}{\bf A}_{x}+2v_{1}{\bf H}\wedge {\bf A}&=&0,\\
{\bf B}_{t}+{\bf B}\wedge {\bf B}_{xx}+u_{2}{\bf B}_{x}+2v_{2}{\bf H}\wedge {\bf B}&=&0, \label{2.2} 
\end{eqnarray} 
or
in matrix form
\begin{eqnarray}
iA_{t}+\frac{1}{2}[A,A_{xx}]+iu_{1}A_{x}+v_{1}[\sigma_{3},A]&=&0,\\
iB_{t}+\frac{1}{2}[B,B_{xx}]+iu_{2}B_{x}+v_{2}[\sigma_{3},B]&=&0, \label{2.2}
\end{eqnarray}
where ${\bf H}=(0,0,1)^{T}$ is the constant magnetic field,  $u_{j}$ and $v_{j}$ are coupling potentials. Here
\begin{eqnarray}
u_{1}&=&i[\left({\bar q}_{2}g_{1}{\bar g}_{3}-q_{2}{\bar g}_{1}g_{3}\right)+\left({\bar q}_{3}g_{1}{\bar g}_{4}-q_{3}{\bar g}_{1}g_{4}\right)],\label{3.666}\\
v_{1}&=&-[|q_{2}|^{2}(\Delta_{1}+|g_{3}|^{2})+|q_{3}|^{2}(\Delta_{1}+|g_{4}|^{2})+q_{2}{\bar q}_{3}g_{3}{\bar g}_{4}+{\bar q}_{2}q_{3}{\bar g}_{3}g_{4}],\label{2.16}\\
u_{2}&=&i[\left({\bar q}_{2}g_{1}{\bar g}_{3}-q_{2}{\bar g}_{1}g_{3}\right)+\left({\bar q}_{3}g_{1}{\bar g}_{4}-q_{3}{\bar g}_{1}g_{4}\right)],\label{3.666}\\
v_{2}&=&-[|q_{2}|^{2}(\Delta_{1}+|g_{3}|^{2})+|q_{3}|^{2}(\Delta_{1}+|g_{4}|^{2})+q_{2}{\bar q}_{3}g_{3}{\bar g}_{4}+{\bar q}_{2}q_{3}{\bar g}_{3}g_{4}],\label{2.16}
\end{eqnarray} 
where  
\begin{eqnarray}
\Delta_{1}&=&|g_{1}|^{2}+|g_{2}|^{2},\label{3.666}\\
\Delta_{2}&=&|g_{1}|^{2}+|g_{3}|^{2},\label{3.666}\\
\Delta&=&|g_{1}|^{2}+|g_{2}|^{2}+|g_{3}|^{2}.\label{3.666}
\end{eqnarray}
In  components,  the coupled M-LIII equation (4.1)-(4.2) reads as
\begin{eqnarray}
iA_{t}^{+}+(A^{+}A_{3xx}-A^{+}_{xx}A_{3})+iu_{1}A^{+}_{x}-2v_{1}A^{+}&=&0,\\
iA_{t}^{-}-(A^{-}A_{3xx}-A^{-}_{xx}A_{3})+iu_{1}A_{x}^{-}+2v_{1}A^{-}&=&0,\\
iA_{3t}+\frac{1}{2}(A^{-}A^{+}_{xx}-A^{-}_{xx}A^{+})+iu_{1}A_{3x}&=&0,\\
iB_{t}^{+}+(B^{+}B_{3xx}-B^{+}_{xx}B_{3})+iu_{2}B_{x}^{+}+2v_{2}B^{+}&=&0, \label{2.2} \\
iB_{t}^{-}-(B^{-}B_{3xx}-B^{-}_{xx}B_{3})+iu_{2}B_{x}^{-}+2v_{2}B^{-}&=&0, \label{2.2} \\
iB_{3t}+\frac{1}{2}(B^{-}B^{+}_{xx}-B^{-}_{xx}B^{+})+iu_{2}B_{3x}&=&0, \label{2.2} 
\end{eqnarray}
or
\begin{eqnarray}
A_{1t}+A_{2}A_{3xx}-A_{2xx}A_{3}+u_{1}A_{1x}-2v_{1}A_{2}&=&0,\\
A_{2t}+A_{3}A_{1xx}-A_{3xx}A_{1}+u_{1}A_{2x}-2v_{1}A_{1}&=&0,\\
A_{3t}+A_{1}A_{2xx}-A_{1xx}A_{2}+u_{1}A_{3x}&=&0,\\
B_{1t}+B_{2}B_{3xx}-B_{2xx}B_{3}+u_{2}B_{1x}-2v_{2}B_{2}&=&0,\\
B_{2t}+B_{3}B_{1xx}-B_{3xx}B_{1}+u_{2}B_{2x}-2v_{2}B_{1}&=&0,\\
B_{3t}+B_{1}B_{2xx}-B_{1xx}B_{2}+u_{2}B_{3x}&=&0. \label{2.2} 
\end{eqnarray}

\subsection{The Lakshmanan equivalent counterpart }

In this subsection we present the Lakshmanan equivalent counterpart of the 2-layer M-LIII equation (2.1)-(2.2)
Now we consider  two interacting 3-dimensional curves in  $R^{n}$. These curves are given by the  following two basic vectors 
${\bf e}_{k}$ and ${\bf l}_{k}$. The  motion of these   curves is defined by the following  
equations 
\begin{eqnarray}
\left ( \begin{array}{ccc}
{\bf  e}_{1} \\
{\bf  e}_{2} \\
{\bf  e}_{3}
\end{array} \right)_{x} = C
\left ( \begin{array}{ccc}
{\bf  e}_{1} \\
{\bf  e}_{2} \\
{\bf  e}_{3}
\end{array} \right),\quad
\left ( \begin{array}{ccc}
{\bf  e}_{1} \\
{\bf  e}_{2} \\
{\bf  e}_{3}
\end{array} \right)_{t} = D
\left ( \begin{array}{ccc}
{\bf  e}_{1} \\
{\bf  e}_{2} \\
{\bf  e}_{3}
\end{array} \right), \label{2.1} 
\end{eqnarray}
and
\begin{eqnarray}
\left ( \begin{array}{ccc}
{\bf  l}_{1} \\
{\bf  l}_{2} \\
{\bf  l}_{3}
\end{array} \right)_{x} = L
\left ( \begin{array}{ccc}
{\bf  l}_{1} \\
{\bf  l}_{2} \\
{\bf  l}_{3}
\end{array} \right),\quad
\left ( \begin{array}{ccc}
{\bf  l}_{1} \\
{\bf  l}_{2} \\
{\bf  l}_{3}
\end{array} \right)_{t} = N
\left ( \begin{array}{ccc}
{\bf  l}_{1} \\
{\bf  l}_{2} \\
{\bf  l}_{3}
\end{array} \right). \label{2.1} 
\end{eqnarray}
Here ${\bf e}_{1}, {\bf e}_{2}$ and ${\bf e}_{3}$ are the unit tangent, normal 
and binormal vectors respectively to the first curve,  ${\bf l}_{1}, {\bf l}_{2}$ and ${\bf l}_{3}$ are the unit tangent, normal 
and binormal vectors respectively to the second curve,  $x$ is the arclength 
parametrising these both  curves. The matrices $C, D, L, N$ are given by
\begin{eqnarray}
C =
\left ( \begin{array}{ccc}
0   & k_{1}     & 0 \\
-k_{1} & 0     & \tau_{1}  \\
0   & -\tau_{1} & 0
\end{array} \right) ,\quad
G =
\left ( \begin{array}{ccc}
0       & \omega_{3}  & -\omega_{2} \\
-\omega_{3} & 0      & \omega_{1} \\
\omega_{2}  & -\omega_{1} & 0
\end{array} \right),\label{2.2} 
\end{eqnarray}
\begin{eqnarray}
L =
\left ( \begin{array}{ccc}
0   & k_{2}     & 0 \\
-k_{2} & 0     & \tau_{2}  \\
0   & -\tau_{2} & 0
\end{array} \right) ,\quad
N =
\left ( \begin{array}{ccc}
0       & \theta_{3}  & -\theta_{2} \\
-\theta_{3} & 0      & \theta_{1} \\
\theta_{2}  & -\theta_{1} & 0
\end{array} \right).\label{2.2} 
\end{eqnarray}
For the   curvatures and torsions of  curves we obtain
\begin{eqnarray}
k_{1} & = & \sqrt{{\bf e}_{1x}^{2}},\quad \tau_{1}= \frac{{\bf e}_{1}\cdot ({\bf e}_{1x} \wedge {\bf e}_{1xx})}{{\bf e}_{1x}^{2}}, \\
k_{2} & = &\sqrt{ {\bf l}_{1x}^{2}},\quad \tau_{2}= \frac{{\bf l}_{1}\cdot ({\bf l}_{1x} \wedge {\bf l}_{1xx})}{{\bf l}_{1x}^{2}}.        \label{2.3}
\end{eqnarray}
The  equations (3.9) and (3.11) are compatible if 
\begin{eqnarray}
C_t - G_x + [C, G] &=& 0,\label{2.4} \\
L_t - N_x + [L, N] &=& 0.\label{2.4} 
\end{eqnarray}
In elements these equations take the form
 \begin{eqnarray}
k_{1t}    & = & \omega_{3x} + \tau_{1} \omega_2, \label{2.5} \\ 
\tau_{1t}      & = & \omega_{1x} - k_{1}\omega_2, \\ \label{2.6} 
\omega_{2x} & = & \tau_{1} \omega_3-k_{1} \omega_1, \label{2.7} 
\end{eqnarray}
and
 \begin{eqnarray}
k_{2t}    & = & \theta_{3x} + \tau_{2} \theta_2, \label{2.5} \\ 
\tau_{2t}      & = & \theta_{1x} - k_{2}\theta_2, \\ \label{2.6} 
\theta_{2x} & = & \tau_{2} \theta_3-k_{2} \theta_1. \label{2.7} 
\end{eqnarray}
Our next step is  the following identifications:
 \begin{eqnarray}
{\bf A}\equiv {\bf e}_{1}, \quad {\bf B}\equiv {\bf l}_{1}. \label{2.7} 
\end{eqnarray}
We also assume that
\begin{eqnarray}
{\bf F}=F_{1}{\bf e}_{1}+F_{2}{\bf e}_{2}+F_{3}{\bf e}_{3}, \quad {\bf E}=E_{1}{\bf l}_{1}+E_{2}{\bf l}_{2}+E_{3}{\bf l}_{3}, \label{2.7} 
\end{eqnarray}
where
\begin{eqnarray}
{\bf F}=2v_{1}{\bf H}\wedge {\bf A}, \quad {\bf E}=2v_{2}{\bf H}\wedge {\bf B}. \label{2.7} 
\end{eqnarray}
Then we obtain
\begin{eqnarray}
k_{1}^{2} & = & {\bf A}_{x}^{2},\\ 
\tau_{1}&=&  \frac{{\bf A}\cdot ({\bf A}_{x} \wedge {\bf A}_{xx})}{{\bf A}_{x}^{2}}, \\
k_{2}^{2} & = & {\bf B}_{x}^{2},\\
\tau_{2}&=&  \frac{{\bf B}\cdot ({\bf B}_{x} \wedge {\bf B}_{xx})}{ {\bf B}_{x}^{2}},        \label{2.3}
\end{eqnarray}
and
\begin{eqnarray}
\omega_{1} & = & -\frac{k_{1xx}+F_{2}\tau_{1}+F_{3x}}{k_{1}}+(\tau_{1}-u_{1})\tau_{1},\\ 
\omega_{2}&=& k_{1x}+F_{3}, \\
\omega_{3} & = &k_{1}(\tau_{1}-u_{1})-F_{2},\\
\theta_{1} & = & -\frac{k_{2xx}+E_{2}\tau_{2}+E_{3x}}{k_{2}}+(\tau_{2}-u_{2})\tau_{2},\\ 
\theta_{2}&=& k_{2x}+E_{3}, \\
\theta_{3} & = &k_{2}(\tau_{2}-u_{2})-E_{2}.        \label{2.3}
\end{eqnarray}
with
\begin{eqnarray}
F_{1}=E_{1}=0.   \label{2.3}
\end{eqnarray}
We now can  write the equations for $k_{j}$ and $\tau_{j}$. They look like
 \begin{eqnarray}
k_{1t}&=&2k_{1x}\tau_{1}+k_{1}\tau_{1x}-(u_{1}k_{1})_{x}-F_{2x}+F_{3}\tau_{1}, \label{2.5} \\ 
\tau_{1t}&=&\left[-\frac{k_{1xx}+F_{2}\tau_{1}+F_{3x}}{k_{1}}+(\tau_{1}-u_{1})\tau_{1}-\frac{1}{2}k_{1}^{2}\right]_{x}-F_{3}k_{1},  \label{2.7} \\
k_{2t}&=&2k_{2x}\tau_{2}+k_{2}\tau_{2x}-(u_{2}k_{2})_{x}-E_{2x}+E_{3}\tau_{2}, \label{2.5} \\ 
\tau_{2t}&=&\left[-\frac{k_{2xx}+E_{2}\tau_{2}+E_{3x}}{k_{2}}+(\tau_{2}-u_{2})\tau_{2}-\frac{1}{2}k_{2}^{2}\right]_{x}-E_{3}k_{2}. \label{2.7} 
\end{eqnarray}
Let us now  introduce new four  real functions $\alpha_{j}$ and $\beta_{j}$ as
\begin{eqnarray}
\alpha_{1}&=&0.5k_{1}\sqrt{1+ \zeta_{1}},\\ 
\beta_{1}&=&\tau_{1}(1+ \xi_{1}),\\ 
\alpha_{2}&=&0.5k_{2}\sqrt{1+ \zeta_{2}},\\
\beta_{2}&=&\tau_{2}(1+ \xi_{2}), \label{2.2} 
\end{eqnarray}
where
\begin{eqnarray}
\zeta_{1}&=&\frac{2|WA^{-}_{x}-MA^{-}|^{2}}{W^{2}(1+A_{3})^{2}{\bf A}^{2}_{x}}-1,\\ 
\zeta_{2}&=&\frac{2|W[(1+A_{3})(1+B_{3})^{-1}B^{-} ]_{x}-M[(1+A_{3})(1+B_{3})^{-1} B^{-}]|^{2}}{W^{2}(1+A_{3})^{2}{\bf B}^{2}_{x}}-1,\label{2.2} \\
\xi_{1}&=&\frac{\bar{R}_{x}R-\bar{R}R_{x}-4i|R|^{2}\nu_{x}}{2i\alpha_{1}^{2}W^{2}(1+A_{3})^{2}\tau_{1}}-1,\\ 
\xi_{2}&=&\frac{\bar{Z}_{x}Z-\bar{Z}Z_{x}-4i|Z|^{2}\nu_{x}}{2i\alpha_{2}^{2}W^{2}(1+A_{3})^{2}\tau_{2}}-1. \label{2.2} 
\end{eqnarray}
Here 
\begin{eqnarray}
W&=&2+\frac{(1+A_{3})(1-B_{3})}{1+B_{3}}=2+K, \label{2.2} \\
M&=&A_{3x}+\frac{A^{+}A^{-}_{x}}{1+A_{3}}+\frac{A_{3x}(1-B_{3})}{1+B_{3}}+\frac{(1+A_{3})B^{+}B^{-}_{x}}{(1+B_{3})^{2}}-\frac{(1+A_{3})(1-B_{3})B_{3x}}{(1+B_{3})^{2}},\\
R&=&WA^{-}_{x}-MA^{-}, \label{2.2} \\
Z&=&W[(1+A_{3})(1+B_{3})^{-1}B^{-} ]_{x}-M[(1+A_{3})(1+B_{3})^{-1} B^{-}].
\end{eqnarray}
with
\begin{eqnarray}
\nu&=& \partial^{-1}_{x}\left[\frac{A_{1}A_{2x}-A_{1x}A_{2}}{(1+A_{3})W}-\frac{(1+A_{3})(B_{1x}B_{2}-B_{1}B_{2x})}{(1+B_{3})^{2}W}\right].\label{2.2} 
\end{eqnarray}

 We now ready to  write the equations for the functions $\alpha_{i}$ and $\beta_{j}$. They   satisfy the following four equations
\begin{eqnarray}
\alpha_{1t}-2\alpha_{1x}\beta_{1}-\alpha_{1}\beta_{1x}&=&0,\\ 
\beta_{1t}+\left[\frac{\alpha_{1xx}}{\alpha_{1}}-\beta_{1}^{2}+2(\alpha_{1}^{2}+\alpha_{2}^{2})\right]_{x}&=&0, \\ 
\alpha_{2t}-2\alpha_{2x}\beta_{2}-\alpha_{2}\beta_{2x}&=&0,\\ 
\beta_{2t}+\left[\frac{\alpha_{2xx}}{\alpha_{2}}-\beta_{2}^{2}+2(\alpha_{1}^{2}+\alpha_{2}^{2})\right]_{x}&=&0. \label{2.2} 
\end{eqnarray}
Let us now  we introduce new two complex functions as 
\begin{eqnarray}
q_{1}&=&\alpha_{1}e^{-i\partial^{-1}_{x}\beta_{1}},\\ q_{2}&=&\alpha_{2}e^{-i\partial^{-1}_{x}\beta_{2}}. \label{2.2} 
\end{eqnarray}
It is not difficult to verify that these functions satisfy the following Manakov system
\begin{eqnarray}
iq_{1t}+q_{1xx}+2(|q_{1}|^{2}+|q_{2}|^{2})q_{1}&=&0,\label{3.21}\\
 iq_{2t}+q_{2xx}+2(|q_{1}|^{2}+|q_{2}|^{2})q_{2}&=&0.\label{3.23}
\end{eqnarray}
 Thus we have shown that the Manakov system (3.60)-(3.61) is the Lakshmanan equivalent counterpart of the 2-layer M-LIII equation or,  in other terminology, the coupled M-LIII equation (4.1)-(4.2). 

\section{The 3-layer  M-LIII equation}
 
\subsection{The  equation }
 In this paper, we consider three spin vectors ${\bf A}=(A_{1}, A_{2}, A_{3})$, ${\bf B}=(B_{1}, B_{2}, B_{3})$ and  ${\bf C}=(C_{1}, C_{2}, C_{3})$, where ${\bf A}^{2}={\bf B}^{2}={\bf C}^{2}=1$. Let these spin vectors satisfy the  3-layer  M-LIII equation of the form
 \begin{eqnarray}
{\bf A}_{t}+{\bf A}\wedge {\bf A}_{xx}+u_{1}{\bf A}_{x}+2v_{1}{\bf H}\wedge {\bf A}&=&0,\\
{\bf B}_{t}+{\bf B}\wedge {\bf B}_{xx}+u_{2}{\bf B}_{x}+2v_{2}{\bf H}\wedge {\bf B}&=&0, \label{2.2} \\
{\bf C}_{t}+{\bf C}\wedge {\bf C}_{xx}+u_{3}{\bf C}_{x}+2v_{3}{\bf H}\wedge {\bf C}&=&0,
\end{eqnarray} 
or
in matrix form
\begin{eqnarray}
iA_{t}+\frac{1}{2}[A,A_{xx}]+iu_{1}A_{x}+v_{1}[\sigma_{3},A]&=&0,\\
iB_{t}+\frac{1}{2}[B,B_{xx}]+iu_{2}B_{x}+v_{2}[\sigma_{3},B]&=&0, \label{2.2} \\
iC_{t}+\frac{1}{2}[C,C_{xx}]+iu_{3}C_{x}+v_{3}[\sigma_{3},C]&=&0.
\end{eqnarray}
Here $u_{j}$ and $v_{j}$ are coupling potentials and have the forms
\begin{eqnarray}
u_{1}&=&i[\left({\bar q}_{2}g_{1}{\bar g}_{3}-q_{2}{\bar g}_{1}g_{3}\right)+\left({\bar q}_{3}g_{1}{\bar g}_{4}-q_{3}{\bar g}_{1}g_{4}\right)],\label{3.666}\\
v_{1}&=&-[|q_{2}|^{2}(\Delta_{1}+|g_{3}|^{2})+|q_{3}|^{2}(\Delta_{1}+|g_{4}|^{2})+q_{2}{\bar q}_{3}g_{3}{\bar g}_{4}+{\bar q}_{2}q_{3}{\bar g}_{3}g_{4}],\label{2.16}\\
u_{2}&=&i[({\bar q}_{1}{\bar g}_{2}+{\bar q}_{3}{\bar g}_{4})g_{1}-(q_{1}g_{2}+q_{3}g_{4}){\bar q}_{1}],\label{3.666}\\
v_{2}&=&-\frac{2}{\Delta_{2}}[|q_{1}|^{2}(\Delta_{2}+|g_{2}|^{2})+|q_{3}|^{2}(\Delta_{2}+|g_{4}|^{2})+q_{1}{\bar q}_{3}g_{2}{\bar g}_{4}+{\bar q}_{1}q_{3}{\bar g}_{2}g_{4}],\label{2.16}\\
u_{3}&=&\frac{2i}{\Delta_{3}}[({\bar q}_{1}{\bar g}_{2}+{\bar q}_{2}{\bar g}_{3})g_{1}-(q_{1}g_{2}+q_{2} g_{3}){\bar g}_{1}],\label{3.666}\\
v_{3}&=&-\frac{2}{\Delta_{3}}[|q_{1}|^{2}(\Delta_{3}+|g_{2}|^{2})+|q_{2}|^{2}(\Delta_{3}+|g_{3}|^{2})+q_{1}{\bar q}_{2}g_{2}{\bar g}_{3}+{\bar q}_{1}q_{2}{\bar g}_{2}g_{3}],\label{2.16}
\end{eqnarray} 
where
\begin{eqnarray}
\Delta_{1}&=&|g_{1}|^{2}+|g_{2}|^{2},\label{3.666}\\
\Delta_{2}&=&|g_{1}|^{2}+|g_{3}|^{2},\label{3.666}\\
\Delta_{3}&=&|g_{1}|^{2}+|g_{4}|^{2},\label{3.666}\\
\Delta&=&|g_{1}|^{2}+|g_{2}|^{2}+|g_{3}|^{2}+|g_{4}|^{2}.\label{3.666}
\end{eqnarray} 
\subsection{The Lakshmanan equivalent counterpart}

In this subsection we obtain  the Lakshmanan equivalent counterpart of the 3-layer M-LIII equation (2.1)-(2.2). For this purpose we  consider  three interacting 3-dimensional curves in  some Eucledian space. These curves are given by the  following three  vectors 
${\bf e}_{k}$,  ${\bf l}_{k}$ and ${\bf n}_{k}$.  These   vectors are   governed  by the following  
equations 
\begin{eqnarray}
\left ( \begin{array}{ccc}
{\bf  e}_{1} \\
{\bf  e}_{2} \\
{\bf  e}_{3}
\end{array} \right)_{x} = C
\left ( \begin{array}{ccc}
{\bf  e}_{1} \\
{\bf  e}_{2} \\
{\bf  e}_{3}
\end{array} \right),\quad
\left ( \begin{array}{ccc}
{\bf  e}_{1} \\
{\bf  e}_{2} \\
{\bf  e}_{3}
\end{array} \right)_{t} = D
\left ( \begin{array}{ccc}
{\bf  e}_{1} \\
{\bf  e}_{2} \\
{\bf  e}_{3}
\end{array} \right), \label{2.1} 
\end{eqnarray}

\begin{eqnarray}
\left ( \begin{array}{ccc}
{\bf  l}_{1} \\
{\bf  l}_{2} \\
{\bf  l}_{3}
\end{array} \right)_{x} = L
\left ( \begin{array}{ccc}
{\bf  l}_{1} \\
{\bf  l}_{2} \\
{\bf  l}_{3}
\end{array} \right),\quad
\left ( \begin{array}{ccc}
{\bf  l}_{1} \\
{\bf  l}_{2} \\
{\bf  l}_{3}
\end{array} \right)_{t} = N
\left ( \begin{array}{ccc}
{\bf  l}_{1} \\
{\bf  l}_{2} \\
{\bf  l}_{3}
\end{array} \right), \label{2.1} 
\end{eqnarray}
\begin{eqnarray}
\left ( \begin{array}{ccc}
{\bf  n}_{1} \\
{\bf  n}_{2} \\
{\bf  n}_{3}
\end{array} \right)_{x} = M
\left ( \begin{array}{ccc}
{\bf  n}_{1} \\
{\bf  n}_{2} \\
{\bf  n}_{3}
\end{array} \right),\quad
\left ( \begin{array}{ccc}
{\bf  n}_{1} \\
{\bf n}_{2} \\
{\bf  n}_{3}
\end{array} \right)_{t} = J
\left ( \begin{array}{ccc}
{\bf  n}_{1} \\
{\bf  n}_{2} \\
{\bf  n}_{3}
\end{array} \right). \label{2.1} 
\end{eqnarray}
 The matrices $C, D, L, N$ are given by
\begin{eqnarray}
C =
\left ( \begin{array}{ccc}
0   & k_{1}     & 0 \\
-k_{1} & 0     & \tau_{1}  \\
0   & -\tau_{1} & 0
\end{array} \right) ,\quad
G =
\left ( \begin{array}{ccc}
0       & \omega_{3}  & -\omega_{2} \\
-\omega_{3} & 0      & \omega_{1} \\
\omega_{2}  & -\omega_{1} & 0
\end{array} \right),\label{2.2} 
\end{eqnarray}
\begin{eqnarray}
L =
\left ( \begin{array}{ccc}
0   & k_{2}     & 0 \\
-k_{2} & 0     & \tau_{2}  \\
0   & -\tau_{2} & 0
\end{array} \right) ,\quad
N =
\left ( \begin{array}{ccc}
0       & \theta_{3}  & -\theta_{2} \\
-\theta_{3} & 0      & \theta_{1} \\
\theta_{2}  & -\theta_{1} & 0
\end{array} \right),\label{2.2} 
\end{eqnarray}
\begin{eqnarray}
M =
\left ( \begin{array}{ccc}
0   & k_{3}     & 0 \\
-k_{3} & 0     & \tau_{3}  \\
0   & -\tau_{3} & 0
\end{array} \right) ,\quad
J=
\left ( \begin{array}{ccc}
0       & \delta_{3}  & -\delta_{2} \\
-\delta_{3} & 0      & \delta_{1} \\
\delta_{2}  & -\delta_{1} & 0
\end{array} \right).\label{2.2} 
\end{eqnarray}
For the   curvatures and torsions of  curves we obtain
\begin{eqnarray}
k_{1} & = & \sqrt{{\bf e}_{1x}^{2}},\quad \tau_{1}= \frac{{\bf e}_{1}\cdot ({\bf e}_{1x} \wedge {\bf e}_{1xx})}{{\bf e}_{1x}^{2}}, \\
k_{2} & = &\sqrt{ {\bf l}_{1x}^{2}},\quad \tau_{2}= \frac{{\bf l}_{1}\cdot ({\bf l}_{1x} \wedge {\bf l}_{1xx})}{{\bf l}_{1x}^{2}},        \label{2.3}\\
k_{3} & = &\sqrt{ {\bf n}_{1x}^{2}},\quad \tau_{3}= \frac{{\bf n}_{1}\cdot ({\bf n}_{1x} \wedge {\bf n}_{1xx})}{{\bf n}_{1x}^{2}}.        \label{2.3}
\end{eqnarray}
The  equations (3.9) and (3.11) are compatible if 
\begin{eqnarray}
C_t - G_x + [C, G] &=& 0,\label{2.4} \\
L_t - N_x + [L, N] &=& 0,\label{2.4} \\
M_t - J_x + [M, J] &=& 0.\label{2.4} 
\end{eqnarray}
In elements these equations take the form
 \begin{eqnarray}
k_{1t}    & = & \omega_{3x} + \tau_{1} \omega_2, \label{2.5} \\ 
\tau_{1t}      & = & \omega_{1x} - k_{1}\omega_2, \\ \label{2.6} 
\omega_{2x} & = & \tau_{1} \omega_3-k_{1} \omega_1 \label{2.7} 
\end{eqnarray}
\begin{eqnarray}
k_{2t}    & = & \theta_{3x} + \tau_{2} \theta_2, \label{2.5} \\ 
\tau_{2t}      & = & \theta_{1x} - k_{2}\theta_2, \\ \label{2.6} 
\theta_{2x} & = & \tau_{2} \theta_3-k_{2} \theta_1. \label{2.7} 
\end{eqnarray}
and
 \begin{eqnarray}
k_{3t}    & = & \delta_{3x} + \tau_{2} \delta_2, \label{2.5} \\ 
\tau_{3t}      & = & \delta_{1x} - k_{3}\delta_2, \\ \label{2.6} 
\delta_{2x} & = & \tau_{3} \delta_3-k_{3} \delta_1. \label{2.7} 
\end{eqnarray}
As in the previous section we assume   the following identifications:
 \begin{eqnarray}
{\bf A}\equiv {\bf e}_{1}, \quad {\bf B}\equiv {\bf l}_{1},  \quad {\bf C}\equiv {\bf n}_{1}. \label{2.7} 
\end{eqnarray}
We also assume that
\begin{eqnarray}
{\bf F}=F_{1}{\bf e}_{1}+F_{2}{\bf e}_{2}+F_{3}{\bf e}_{3}, \quad {\bf E}=E_{1}{\bf l}_{1}+E_{2}{\bf l}_{2}+E_{3}{\bf l}_{3}, \quad {\bf P}=P_{1}{\bf n}_{1}+P_{2}{\bf n}_{2}+P_{3}{\bf n}_{3},  \label{2.7} 
\end{eqnarray}
where
\begin{eqnarray}
{\bf F}=2v_{1}{\bf H}\wedge {\bf A}, \quad {\bf E}=2v_{2}{\bf H}\wedge {\bf B}, \quad {\bf P}=2v_{3}{\bf H}\wedge {\bf C}. \label{2.7} 
\end{eqnarray}
Then we obtain
\begin{eqnarray}
k_{1}^{2} & = & {\bf A}_{x}^{2},\\ 
\tau_{1}&=&  \frac{{\bf A}\cdot ({\bf A}_{x} \wedge {\bf A}_{xx})}{{\bf A}_{x}^{2}}, \\
k_{2}^{2} & = & {\bf B}_{x}^{2},\\
\tau_{2}&=&  \frac{{\bf B}\cdot ({\bf B}_{x} \wedge {\bf B}_{xx})}{ {\bf B}_{x}^{2}},        \label{2.3}\\
k_{3}^{2} & = & {\bf C}_{x}^{2},\\
\tau_{3}&=&  \frac{{\bf C}\cdot ({\bf C}_{x} \wedge {\bf C}_{xx})}{ {\bf C}_{x}^{2}},        \label{2.3}
\end{eqnarray}
and
\begin{eqnarray}
\omega_{1} & = & -\frac{k_{1xx}+F_{2}\tau_{1}+F_{3x}}{k_{1}}+(\tau_{1}-u_{1})\tau_{1},\\ 
\omega_{2}&=& k_{1x}+F_{3}, \\
\omega_{3} & = &k_{1}(\tau_{1}-u_{1})-F_{2},\\
\theta_{1} & = & -\frac{k_{2xx}+E_{2}\tau_{2}+E_{3x}}{k_{2}}+(\tau_{2}-u_{2})\tau_{2},\\ 
\theta_{2}&=& k_{2x}+E_{3}, \\
\theta_{3} & = &k_{2}(\tau_{2}-u_{2})-E_{2}, \\
\delta_{1} & = & -\frac{k_{3xx}+P_{2}\tau_{3}+P_{3x}}{k_{3}}+(\tau_{3}-u_{3})\tau_{3},\\ 
\delta_{2}&=& k_{3x}+P_{3}, \\
\delta_{3} & = &k_{3}(\tau_{3}-u_{3})-P_{2},       \label{2.3}
\end{eqnarray}
with
\begin{eqnarray}
F_{1}=E_{1}=P_{1}=0.   \label{2.3}
\end{eqnarray}
As a result, we obtain  the following equations for $k_{j}$ and $\tau_{j}$. They read as 
 \begin{eqnarray}
k_{1t}&=&2k_{1x}\tau_{1}+k_{1}\tau_{1x}-(u_{1}k_{1})_{x}-F_{2x}+F_{3}\tau_{1}, \label{2.5} \\ 
\tau_{1t}&=&\left[-\frac{k_{1xx}+F_{2}\tau_{1}+F_{3x}}{k_{1}}+(\tau_{1}-u_{1})\tau_{1}-\frac{1}{2}k_{1}^{2}\right]_{x}-F_{3}k_{1},  \label{2.7} \\
k_{2t}&=&2k_{2x}\tau_{2}+k_{2}\tau_{2x}-(u_{2}k_{2})_{x}-E_{2x}+E_{3}\tau_{2}, \label{2.5} \\ 
\tau_{2t}&=&\left[-\frac{k_{2xx}+E_{2}\tau_{2}+E_{3x}}{k_{2}}+(\tau_{2}-u_{2})\tau_{2}-\frac{1}{2}k_{2}^{2}\right]_{x}-E_{3}k_{2}, \label{2.7} \\
k_{3t}&=&2k_{3x}\tau_{3}+k_{3}\tau_{3x}-(u_{3}k_{3})_{x}-P_{2x}+P_{3}\tau_{3}, \label{2.5} \\ 
\tau_{3t}&=&\left[-\frac{k_{3xx}+P_{2}\tau_{3}+P_{3x}}{k_{3}}+(\tau_{3}-u_{3})\tau_{3}-\frac{1}{2}k_{3}^{2}\right]_{x}-P_{3}k_{3}. \label{2.7} 
\end{eqnarray}
According to our approach (see e.g. the refs. \cite{akbota1}-\cite{akbota7}),  we  now  introduce the following new  functions $\alpha_{j}$ and $\beta_{j}$ as
\begin{eqnarray}
\alpha_{1}&=&0.5k_{1}\sqrt{1+ \zeta_{1}},\\ 
\beta_{1}&=&\tau_{1}(1+ \xi_{1}),\\ 
\alpha_{2}&=&0.5k_{2}\sqrt{1+ \zeta_{2}},\\
\beta_{2}&=&\tau_{2}(1+ \xi_{2}), \label{2.2} \\
\alpha_{3}&=&0.5k_{3}\sqrt{1+ \zeta_{3}},\\
\beta_{3}&=&\tau_{3}(1+ \xi_{3}). \label{2.2} 
\end{eqnarray}
where $\zeta_{j}$ and $ \xi_{j}$ are some real functions.
 We  now can show that the equations for the functions $\alpha_{i}$ and $\beta_{j}$ read as
\begin{eqnarray}
\alpha_{1t}-2\alpha_{1x}\beta_{1}-\alpha_{1}\beta_{1x}&=&0,\\ 
\beta_{1t}+\left[\frac{\alpha_{1xx}}{\alpha_{1}}-\beta_{1}^{2}+2(\alpha_{1}^{2}+\alpha_{2}^{2}+\alpha_{3}^{2})\right]_{x}&=&0, \\ 
\alpha_{2t}-2\alpha_{2x}\beta_{2}-\alpha_{2}\beta_{2x}&=&0,\\ 
\beta_{2t}+\left[\frac{\alpha_{2xx}}{\alpha_{2}}-\beta_{2}^{2}+2(\alpha_{1}^{2}+\alpha_{2}^{2}+\alpha_{3}^{2})\right]_{x}&=&0,\label{2.2} \\
\alpha_{3t}-2\alpha_{3x}\beta_{3}-\alpha_{3}\beta_{3x}&=&0,\\ 
\beta_{3t}+\left[\frac{\alpha_{3xx}}{\alpha_{3}}-\beta_{3}^{2}+2(\alpha_{1}^{2}+\alpha_{2}^{2}+\alpha_{3}^{2})\right]_{x}&=&0. \label{2.2} 
\end{eqnarray}
Next we introduce new three  complex functions $q_{j}$  as 
\begin{eqnarray}
q_{1}&=&\alpha_{1}e^{-i\partial^{-1}_{x}\beta_{1}},\\ q_{2}&=&\alpha_{2}e^{-i\partial^{-1}_{x}\beta_{2}}, \label{2.2} \\
q_{3}&=&\alpha_{3}e^{-i\partial^{-1}_{x}\beta_{3}}. \label{2.2}
\end{eqnarray}
The straight calculation shows that these  functions obey  the following vector NLS equation
\begin{eqnarray}
iq_{1t}+q_{1xx}+2(|q_{1}|^{2}+|q_{2}|^{2}+|q_{3}|^{2})q_{1}&=&0,\label{3.21}\\
 iq_{2t}+q_{2xx}+2(|q_{1}|^{2}+|q_{2}|^{2}+|q_{3}|^{2})q_{2}&=&0,\label{3.23}\\
 iq_{3t}+q_{3xx}+2(|q_{1}|^{2}+|q_{2}|^{2}+|q_{3}|^{2})q_{3}&=&0.\label{3.23}
\end{eqnarray}
 Thus we have proved  that the 3-component NLS equation (5.78)-(5.80)  is the Lakshmanan equivalent counterpart of the 3-layer M-LIII equation  (2.1)-(2.2). 


\section{The 4-layer  M-LIII equation}
\subsection{The  equation }
 In this section we  study the 4-layer  M-LIII equation. It is given  as
 \begin{eqnarray}
{\bf A}_{t}+{\bf A}\wedge {\bf A}_{xx}+u_{1}{\bf A}_{x}+2v_{1}{\bf H}\wedge {\bf A}&=&0,\\
{\bf B}_{t}+{\bf B}\wedge {\bf B}_{xx}+u_{2}{\bf B}_{x}+2v_{2}{\bf H}\wedge {\bf B}&=&0, \label{2.2} \\
{\bf C}_{t}+{\bf C}\wedge {\bf C}_{xx}+u_{3}{\bf C}_{x}+2v_{3}{\bf H}\wedge {\bf C}&=&0,\\
{\bf D}_{t}+{\bf D}\wedge {\bf D}_{xx}+u_{4}{\bf D}_{x}+2v_{4}{\bf H}\wedge {\bf D}&=&0. \label{2.2} \\
\end{eqnarray} 
In  matrix form, this equation takes the form
\begin{eqnarray}
iA_{t}+\frac{1}{2}[A,A_{xx}]+iu_{1}A_{x}+v_{1}[\sigma_{3},A]&=&0,\\
iB_{t}+\frac{1}{2}[B,B_{xx}]+iu_{2}B_{x}+v_{2}[\sigma_{3},B]&=&0, \label{2.2} \\
iC_{t}+\frac{1}{2}[C,C_{xx}]+iu_{3}C_{x}+v_{3}[\sigma_{3},C]&=&0,\\
iD_{t}+\frac{1}{2}[D,D_{xx}]+iu_{4}D_{x}+v_{4}[\sigma_{3},D]&=&0, \label{2.2} 
\end{eqnarray}
where   ${\bf A}=(A_{1}, A_{2}, A_{3})$, ${\bf B}=(B_{1}, B_{2}, B_{3})$,  ${\bf C}=(C_{1}, C_{2}, C_{3})$, ${\bf D}=(D_{1}, D_{2}, D_{3})$ and ${\bf A}^{2}={\bf B}^{2}={\bf C}^{2}={\bf D}^{2}=1$. Here $u_{j}$ and $v_{j}$ are coupling potentials.
\subsection{The Lakshmanan equivalent counterpart}

The similar  algebra as in the previous sections shows that the Lakshmanan equivalent counterpart of the 4-layer M-LIII equation (2.1)-(2.2) reads as
\begin{eqnarray}
iq_{1t}+q_{1xx}+2(|q_{1}|^{2}+|q_{2}|^{2}+|q_{3}|^{2}+|q_{4}|^{2})q_{1}&=&0,\label{3.21}\\
iq_{2t}+q_{2xx}+2(|q_{1}|^{2}+|q_{2}|^{2}+|q_{3}|^{2}+|q_{4}|^{2})q_{2}&=&0.\label{3.23}\\
iq_{3t}+q_{3xx}+2(|q_{1}|^{2}+|q_{2}|^{2}+|q_{3}|^{2}+|q_{4}|^{2})q_{3}&=&0.\label{3.23}\\
iq_{4t}+q_{4xx}+2(|q_{1}|^{2}+|q_{2}|^{2}+|q_{3}|^{2}+|q_{4}|^{2})q_{4}&=&0.\label{3.23}
\end{eqnarray}

\section{The N-layer  M-LIII equation}
 \subsection{The  equation }
 In this section we present  the $N$-layer  M-LIII equation. This equation can be written  as
 \begin{eqnarray}
{\bf A}^{(1)}_{t}+{\bf A}^{(1)}\wedge {\bf A}^{(1)}_{xx}+u_{1}{\bf A}^{(1)}_{x}+2v_{1}{\bf H}\wedge {\bf A}^{(1)}&=&0,\\
{\bf A}^{(2)}_{t}+{\bf A}^{(2)}\wedge {\bf A}^{(2)}_{xx}+u_{2}{\bf A}^{(2)}_{x}+2v_{2}{\bf H}\wedge {\bf A}^{(2)}&=&0,\\
\vdots\\
{\bf A}^{(N)}_{t}+{\bf A}^{(N)}\wedge {\bf A}^{(N)}_{xx}+u_{N}{\bf A}^{(N)}_{x}+2v_{N}{\bf H}\wedge {\bf A}^{(N)}&=&0.
\end{eqnarray} 
Let us rewrite this equation in the matrix form as
\begin{eqnarray}
iA_{t}^{(1)}+\frac{1}{2}[A^{(1)},A_{xx}^{(1)}]+iu_{1}A_{x}^{(1)}+v_{1}[\sigma_{3},A^{(1)}]&=&0,\\
iA_{t}^{(2)}+\frac{1}{2}[A^{(2)},A_{xx}^{(2)}]+iu_{2}A_{x}^{(2)}+v_{2}[\sigma_{3},A^{(2)}]&=&0,\\
\vdots\\
iA_{t}^{(N)}+\frac{1}{2}[A^{(N)},A_{xx}^{(N)}]+iu_{N}A_{x}^{(N)}+v_{N}[\sigma_{3},A^{(N)}]&=&0,\end{eqnarray}
where   ${\bf A}^{(j)}=(A_{1}^{(j)}, A_{2}^{(j)}, A_{3}^{(j)})$ and  ${\bf A}^{(j)2}=1$ with 
\begin{eqnarray}
A^{(j)}=\begin{pmatrix} A_{3}^{(j)}&A^{(j)-}\\ 
A^{(j)+}&-A_{3}^{(j)}\end{pmatrix}, \quad A^{(j)2}=I, \quad A^{(j)\pm}=A_{1}^{(j)}\pm i A_{2}^{(j)}, \label{2.2} 
\end{eqnarray}
and $u_{j}$ and $v_{j}$ are some coupling potentials.
\subsection{The Lakshmanan equivalent counterpart}

We now can write  the Lakshmanan equivalent counterpart of the N-layer M-LIII equation (7.1)-(7.4). It is the following vector NLSE
\begin{eqnarray}
iq_{1t}+q_{1xx}+2(|q_{1}|^{2}+|q_{2}|^{2}+\cdots +|q_{N}|^{2})q_{1}&=&0,\label{3.21}\\
iq_{2t}+q_{2xx}+2(|q_{1}|^{2}+|q_{2}|^{2}+\cdots +|q_{N}|^{2})q_{2}&=&0,\label{3.23}\\
\vdots\\
iq_{Nt}+q_{Nxx}+2(|q_{1}|^{2}+|q_{2}|^{2}+\cdots +|q_{N}|^{2})q_{N}&=&0. \label{3.23}
\end{eqnarray}

 \section{Hamiltonian structure}
  In this section we briefly present main elements of the Hamiltonian structure of the multilayer M-LIII equation. As example, here we consider only $N=3$ case that is the 3-layer M-LIII equation (5.1)-(5.3). 
Let us start from quantum Heisenberg magnetic model for the 3-layer case. We write it as
\begin{eqnarray}
\hat{H}=J\sum ( \hat{A}_{i}\hat{A}_{i+1}+\hat{B}_{i}\hat{B}_{i}+\hat{C}_{i}\hat{C}_{i})+\hat{H}_{int}. \label{2.2} 
\end{eqnarray}
In classical limit, this model takes the form
\begin{eqnarray}
H=J\sum (A_{i}A_{i+1}+B_{i}B_{i}+C_{i}C_{i})+H_{int}. \label{2.2} 
\end{eqnarray}
The continuum limit of this model looks like
   \begin{eqnarray}
H=H_{a}+H_{b}+H_{c}+H_{int}, \label{2.2} 
\end{eqnarray}
where
  \begin{eqnarray}
H_{a}=0.5\int {\bf A}_{x}^{2}dx, \quad  H_{b}=0.5\int {\bf B}_{x}^{2}dx,\quad  H_{c}=0.5\int {\bf C}_{x}^{2}dx, H_{int}=0.5\int h_{int}dx, \label{2.2} 
\end{eqnarray}
with $h_{int}=h_{int}(A_{j}, B_{j}, C_{j}, A_{jx}, B_{jx}, C_{jx})$. The Hamilton form of the 3-layer M-LIII equation (5.1)-(5.3) is given by
 \begin{eqnarray}
A_{it}&=&\{H,A_{i}\}=\epsilon_{ijk}\frac{\delta  H}{\delta A_{j}}A_{k}, \label{2.2} \\
B_{it}&=&\{H,B_{i}\}=\epsilon_{ijk}\frac{\delta  H}{\delta B_{j}}B_{k}, \label{2.2}\\
C_{it}&=&\{H,C_{i}\}=\epsilon_{ijk}\frac{\delta  H}{\delta C_{j}}C_{k}. \label{2.2}
\end{eqnarray}
Here the Poisson bracket is given by
\begin{eqnarray}
\{P,Q\}=\epsilon_{ijk}\int\left[\frac{\delta  P}{\delta A_{i}}\frac{\delta  Q}{\delta A_{j}}A_{k}+\frac{\delta  P}{\delta B_{i}}\frac{\delta  Q}{\delta B_{j}}B_{k}+\frac{\delta  P}{\delta C_{i}}\frac{\delta  Q}{\delta C_{j}}C_{k}\right]dx. \label{2.2}
\end{eqnarray}
Hence for the components of the spin vectors we obtain
\begin{eqnarray}
\{A_{i},A_{j}\}&=&\epsilon_{ijk}A_{k}\delta(x-y), \label{2.2}\\
\{B_{i},B_{j}\}&=&\epsilon_{ijk}B_{k}\delta(x-y), \label{2.2}\\
\{C_{i},C_{j}\}&=&\epsilon_{ijk}C_{k}\delta(x-y). \label{2.2}
\end{eqnarray}
and 
\begin{eqnarray}
\{A_{i},B_{j}\}=\{A_{i},B_{j}\}=\{A_{i},C_{j}\}=0. \label{2.2}
\end{eqnarray}
 \section{The $\Gamma$ -- spin system}
 
In the previous section we have shown that between the multilayer M-LIII equation and the vector NLSE takes place the Lakshmanan equivalence. But it is well known that there exists  another "spin system", namely,  the $\Gamma$-spin system which is also (gauge) equivalent to the vector NLSE. The $\Gamma$-spin system reads as\begin{eqnarray}
i\Gamma_{t}+\frac{1}{2}[\Gamma, \Gamma_{xx}]=0, \label{2.2} 
\end{eqnarray}
where
\begin{eqnarray}
\Gamma =(\Gamma_{ij})\in su(N+1).\label{2.2} 
\end{eqnarray}

\section{The relation between solutions of the multilayer M-LIII equation and the $\Gamma$-spin system}
It is natural that the solutions of the multilayer M-LIII equation (4.3)-(4.4) and the $\Gamma$-spin system (9.1) is related to each other by some exact transformations. As example, here we present these transformations for the 2-layer M-LIII equation that is for $N= 2$ case. 
\subsection{Direct  M-transformation}

Let  ${\bf A}$ and ${\bf B}$ be the solution of the 2-layer M-LIII equation (4.3)-(4.4). Then, according to the direct M-transformation \cite{akbota1}-\cite{akbota7}, the solutions of  the $\Gamma$-spin system (9.1)  are expressed as
\begin{eqnarray}
\Gamma =\frac{1}{2+K}
\left ( \begin{array}{ccc}
2A_{3}-K   & 2A^{-}     & \frac{2(1+A_{3})B^{-}}{1+B_{3}} \\
2A^{+} &   -(2A_{3}+K)&\frac{2A^{+}B^{-}}{1+B_{3}} \\
\frac{2(1+A_{3})B^{+}}{1+B_{3}}   & \frac{2A^{-}B^{+}}{1+B_{3}} & K-2
\end{array} \right),\label{2.2} 
\end{eqnarray}
where
\begin{eqnarray}
K= \frac{(1+A_{3})(1-B_{3})}{1+B_{3}}.\label{2.2} 
\end{eqnarray}

\subsection{Inverse M-transformation}
We now consider  the inverse M-transformation. Let $\Gamma_{ij}$ be solutions of the $\Gamma$ -- spin system (9.1). Then the  solutions of the coupled or (that same) 2-layer  M-LIII equation (4.3)-(4.4) are  defined  as
\begin{eqnarray}
A &=&\frac{1}{1-\Gamma_{33}}
\left ( \begin{array}{cc}
\Gamma_{11}-\Gamma_{22}   & 2\Gamma_{12} \\
2\Gamma_{21} &   \Gamma_{22}-\Gamma_{11}\end{array}\right),\label{2.2} \\
B&=&\frac{1}{1-\Gamma_{22}}
\left ( \begin{array}{cc}
\Gamma_{11}-\Gamma_{33}   & 2\Gamma_{13} \\
2\Gamma_{31} &   \Gamma_{33}-\Gamma_{11}\end{array}\right),\label{2.2} 
\end{eqnarray}
 where $\Gamma_{ij}$ are given by the formulas (9.2). Similarly, we find the direct and inverse M-transformations for $N\neq 2$ cases.

 \section{Geometric flows of immersed surfaces}
 Let ${\bf r}_{j}={\bf r}_{j}(x,t)$ be the position vector of the immersed $j$-th surface in the some Euclidean  space. Such surfaces are  given by the following set of first fundamental forms
 \begin{eqnarray}
I_{1}&=&dx^{2}+2{\bf r}_{1x}\cdot{\bf r}_{1t}dxdt+{\bf r}_{1t}^{2}dt^{2},\label{2.2} \\
I_{2}&=&dx^{2}+2{\bf r}_{2x}\cdot{\bf r}_{2t}dxdt+{\bf r}_{2t}^{2}dt^{2},\label{2.2} \\
\vdots\\
I_{N}&=&dx^{2}+2{\bf r}_{Nx}\cdot{\bf r}_{Nt}dxdt+{\bf r}_{Nt}^{2}dt^{2},\label{2.2} 
\end{eqnarray}
where we assumed that ${\bf r}_{x}^{2}=1$. We write the set of the second fundamental forms as
\begin{eqnarray}
II_{1}&=&L_{1}dx^{2}+2M_{1}dxdt+N_{1}dt^{2},\label{2.2} \\
II_{2}&=&L_{2}dx^{2}+2M_{2}dxdt+N_{2}dt^{2},\label{2.2} \\
\vdots\\
II_{N}&=&L_{N}dx^{2}+2M_{N}dxdt+N_{N}dt^{2}.\label{2.2} 
\end{eqnarray}
Finally, we can also write the  third fundamental forms. As it is well known, the  third fundamental forms can be written in terms of the first and second forms as
\begin{eqnarray}
III_{1}&=&2H_{1} II_{1} -K_{1} I_{1},\label{2.2} \\
III_{2}&=&2H_{2} II_{2} -K_{2} I_{2},\label{2.2} \\
\vdots\\
III_{N}&=&2H_{N} II_{N} -K_{N} I_{N},\label{2.2} 
\end{eqnarray}
where $H_{j}$ and $K_{j}$ are  the mean curvature and the Gaussian curvature of the $j$-th surface respectively.

\section{Geometric flows of curves}

In this section we want to present another but the equivalent approach to derive the Lakshmanan equivalent of the 1-layer M-LIII equation (3.5). For this purpose we consider the curve which is  given by
\begin{eqnarray}
{\bf r}_t=a{\bf e}_{1}+b{\bf e}_{2}+c{\bf e}_{3},
\end{eqnarray}
where $\bf{e}_{1}$, $\bf{e}_{2}$ and $\bf{e}_{3}$ denote the tangent, normal and binormal vectors of the curve,
respectively. 
The velocities $a$, $b$ and $c$ depend on the $\kappa$  and $\tau$ as well as their
derivatives with respect to arclength parameter $x$.
The arclength parameter $x$ is defined
implicitly by $ds=hdp$, $h=|{\bf r}(p)|$, where $p$ is a~free parameter and is independent of
time.
From  the flow (12.1), the time evolutions of the vectors ${\bf e}_{j}$  are given
by
\begin{eqnarray}
\dot{\bf e}_{3}&=&\left( a_{x}-\tau b
+\kappa c\right){\bf e}_{1}+\left(b_{x}+\tau a\right){\bf e}_{2},\\
\dot{\bf e}_{1}&=&-\left(a_{x}-\tau b+\kappa c){\bf e}_{3}+
\left[\frac{1}{\kappa}(b_{x}+\tau a\right)_{x}+\frac{\tau}{\kappa}(a_{x}
-\tau b+\kappa c)\right]{\bf e}_{2},\\
\dot{\bf e}_{2}&=&-\left(b_{x}+\tau a\right){\bf e}_{3}-
\left[\frac{1}{\kappa} (b_{x}+\tau a)_{x}+\frac{\tau}{\kappa}(a_{x}-\tau b+\kappa c)\right]{\bf e}_{1},\\
\dot{h}&=&2h(c_{x}-\kappa a)\label{12.5}
\end{eqnarray}
and
\begin{eqnarray}
\tau_{t}&=&\left[\frac{1}
{\kappa}\left( b_{x}+\tau a\right)+\frac{\tau}{\kappa}\left(a_{x}-
\tau b\right)+\tau\int^x\kappa a{\rm d} x'\right]+\kappa\tau a+\kappa b_{x},\\
\kappa_{t}&=&a_{xx}
+\big(\kappa^2-\tau^2\big)a+\kappa_{x}\int^x\kappa a{\rm d} x'
-2\tau b_{x}-\tau_{x}b.\label{12.6}
\end{eqnarray}

Assuming that the flow is intrinsic, namely that the arclength does not depend on time, it implies from (12.5) that
\begin{gather}\label{W-U}
c_s=\kappa a.
\end{gather}
Let $a=0$, $b=\kappa$, where $\kappa$ is a~real function, then (12.8)
implies that $c=c_1$, where $c_1$ is a~constant.
Let  $c_1=0$, then the set of equations (12.2)-(12.4)  takes the form
\begin{eqnarray}
\dot{\bf e}_{3}&=&-\tau\kappa{\bf e}_{1}+\kappa_{x}{\bf e}_{2},\nonumber\\
\dot{\bf e}_{1}&=&\tau\kappa{\bf e}_{3}+
\left[\frac{\kappa_{xx}}{\kappa}-\tau^{2}\right]{\bf e}_{2},\nonumber\\
\dot{\bf e}_{2}&=&-\kappa_{x}{\bf e}_{3}-\left[\frac{\kappa_{xx}}{\kappa}-\tau^{2}\right]{\bf e}_{1}.\label{12.7}
\end{eqnarray}

\section{Integrable  filament equations of interacting vortices}
Let us consider the following r-form of the multilayer M-LIII equation, shortly, the multilayer r-M-LIII equation
\begin{eqnarray}\label{S-Flow}
{\bf r}_{jt}={\bf r}_{jx}\wedge {\bf r}_{jxx}+u_{j}{\bf r}_{jx}+2v_{j}{\bf H}\wedge {\bf r}_{j}+{\bf L}_{j},
\end{eqnarray}
where $j=1, 2, \cdots,  N$ and 
\begin{eqnarray}\label{S-Flow}
{\bf L}_{j}=-\partial_{x}^{-1}[u_{jx}{\bf r}_{jx}+2v_{jx}{\bf H}\wedge {\bf r}_{j}].
\end{eqnarray} 
This closed set of equations is integrable. It describes the (integrable) interaction of $N$  vortices.  Indeed, this system is  the  closed set of the  filament equations for interacting $N$ vortices. 
In the case $u_{j}=v_{j}=0$,  we obtain the following uncoupled (noninteracting)  $N$ vortices filament equations
\begin{eqnarray}\label{S-Flow}
{\bf r}_{jt}={\bf r}_{jx}\wedge {\bf r}_{jxx}.
\end{eqnarray}

 \section{Conclusion}
 
 In this paper, we have shown that the  $N$-layer M-LIII equation can be related with the geometric flows of interacting curves and surfaces in some space $R^{n}$. Then we have found the Lakshmanan equivalent counterparts of the   $N$-layer M-LIII equations. After some algebra we have proved that these counterparts in fact are  the vector NLSE. On the other hand, it is well-known that the vector NLSE  is equivalent to the $\Gamma$-spin system. Also,  we have presented the transformations which give the relation between solutions of the $\Gamma$-spin system and the multilayer  M-LIII equation. It is interesting to understand   the role of the  constant magnetic field ${\bf H}$. It seems that this constant magnetic vector plays an important role in our construction of integrable multilayer spin systems and in nonlinear dynamics of multilayer magnetic systems.  

 \end{document}